\documentclass[11pt]{article}
\usepackage{blois,epsfig}

\def\Journal#1#2#3#4{{#1} {\bf #2}, #3 (#4)}

\def\NIMA{{\em Nucl. Instrum. Methods} A}

\def\PLB{{\em Phys. Lett.}  B}

\def\be{\begin{equation}}
\def\ee{\end{equation}}
\def\bea{\begin{eqnarray}}
\def\eea{\end{eqnarray}}

\begin{document}
\vspace*{4cm}
\title{CMS EXPERIMENT AT LHC: COMMISSIONING AND EARLY PHYSICS}

\author{ A.N. SAFONOV \\
(for the CMS Collaboration)}

\address{Department of Physics, Texas A\&M University,\\
College Station, Texas 77845, USA}

\maketitle\abstracts{
The CMS collaboration used the past year to greatly improve the level of detector readiness 
for the first collisions data. The acquired operational experience over this year, large gains 
in understanding the detector and improved preparedness for early physics will be instrumental 
in minimizing the time from the first collisions to first LHC physics. The following describes
the status of the CMS experiment and outlines early physics plans with the first LHC 
data.}

\section{Introduction}
With its first collisions expected at the end of 2009, the Large Hadron Collider (LHC) 
will become a new energy frontier in high energy physics. Experimental searches 
at the LHC are widely anticipated to yield major advances in 
particle physics explaining the mechanism of the electroweak symmetry breaking (EWSB) and 
revealing the underlying theory of matter and fundamental interactions. The LHC physics 
program is reach and broad aiming at a wide range of signatures in which new phenomena 
may reveal itself, from Higgs and well studied SM extensions such as Supersymmetry 
(SUSY) and Extra Dimensions to something yet unexpected.

Two flagship multi-purpose LHC experiments, CMS~\cite{CMS} and ATLAS~\cite{ATLAS}, have been 
actively preparing for the first LHC collisions. This 
contribution reports on the status of the CMS detector readiness for data and outlines early 
physics measurements aiming at establishing baseline detector performance. In addition to its 
focus on robust operation and collection of high quality data needed for Higgs and SUSY searches,
CMS will also pursue a number of well motivated new physics scenarios accessible with already 
small amounts of LHC data.

\section{CMS Detector Status and Commissioning}

The Collider Muon Solenoid (CMS) experiment is a multi-purpose particle physics detector 
consisting of several sub-systems providing identification and momentum measurement of
particles produced in the LHC proton-proton collisions. The main sub-systems are 
tracking (pixel and silicon strip detectors), calorimetry, muon chambers, data acquisition
and trigger. 

The construction and integration of the CMS sub-systems has been mostly completed and 
the experiment was ready for data in the Fall of 2008, even collecting data during
the LHC single beam operations. Since the unexpected delay in the LHC schedule associated 
with the October incident, the CMS experiment has embarked on accomplishing many of the 
commissioning tasks typically performed with the early collisions data. To that end, the
CMS has staged a several months long exercise collecting cosmic ray data. During these 
tests, the CMS detector was continuously operating for several months with the fully 
deployed data acquisition and trigger systems. The data collected were used to accomplish 
many calibration and alignment tasks; a fraction of these measurements is described in 
what follows.

\subsection{Commissioning with the 2008 Beam Data}

\begin{figure}
\centerline{
\epsfig{figure=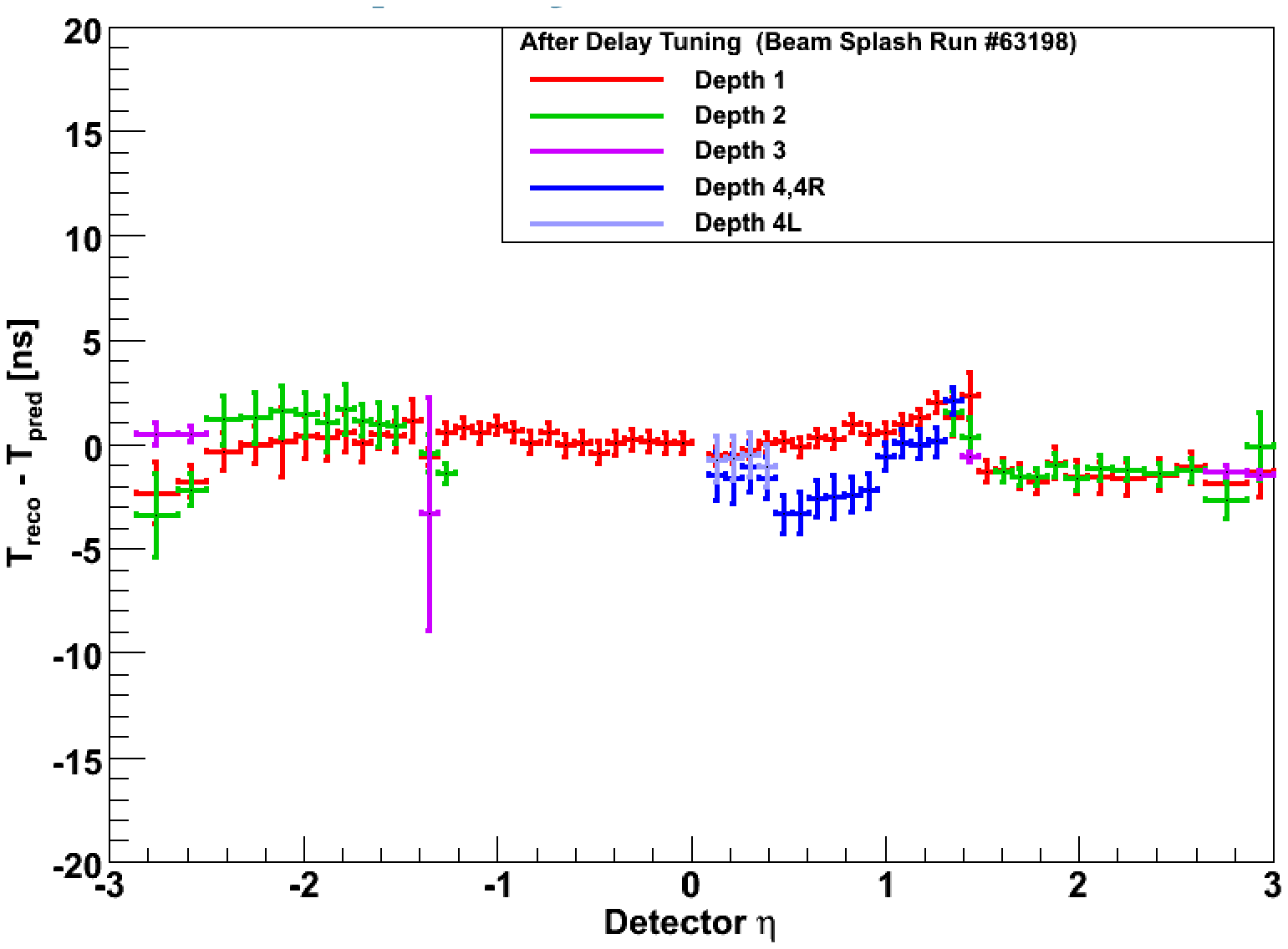,width=0.48\linewidth, angle=0}
\epsfig{figure=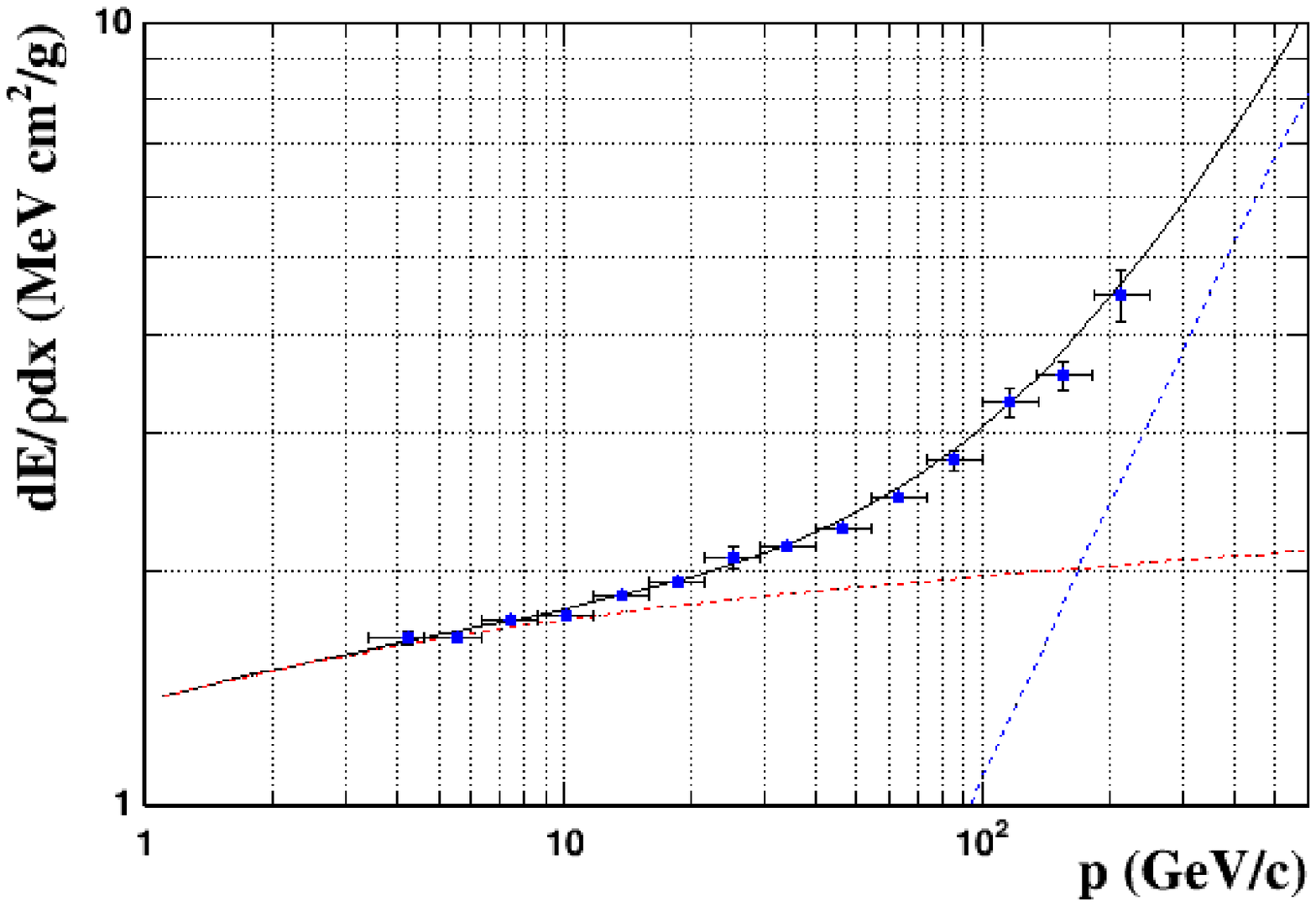,width=0.50\linewidth, angle=0}
}
\caption{(a) Timing in of the hadronic calorimeter using beam splash events from a nine minute long
single beam LHC run (prior to timing corrections, the $\Delta T$ was scattered from $-15$ to $+15$ ns).
(b): Muon stopping power $dE/dx$ measured by the calibrated CMS electromagnetic 
calorimeter as a function of muon momentum compared to the expectation. \label{fig:fig1} }
\end{figure}

In the Fall of 2008, CMS collected data during the LHC single beam running. While not
optimal for ultimate commissioning, these data were used for a number of studies, e.g. 
to time in the calorimeters using spectacular beam splash events that lightened the
entire calorimeter as particles were traversing the detector, see Fig.~\ref{fig:fig1}(a). The single 
beam data was also used to align parts of the CMS Endcap Muon (EMU) system. The EMU system 
consists of a set of cathode-strip chambers arranged in several rings around the beamline.
Because EMU chambers are positioned perpendicular to the beamline, the beam halo muons 
were used to make effectively an X-ray test of the system. The small overlaps of the 
chambers allowed accurate alignment of all EMU rings that had all chambers operational 
during the 9 minute long single beam run. This measurement has demonstrated that with 
just a few minutes of beam data, the EMU system can be aligned to the design accuracy of 
$\simeq$250 $\mu$m.

\subsection{Commissioning with the 2008 Cosmic Ray Data}

Apart from providing invaluable operational experience of continuously 
running a large and highly complex system, the cosmic ray exercise provided 
a wealth of cosmic ray data. Analysis of the data allowed a large improvement 
in understanding the detector alignment, calibration as well as tuning and 
validating reconstruction algorithms.

Figure~\ref{fig:fig1}(b) shows the measurement of energy deposition by cosmic ray
muons in the calibrated CMS electromagnetic calorimeter as a function of muon momentum.
This measurement is compared to the expectation~\cite{PDG} for muon stopping power $dE/dx$ in 
$PbWO_4$ showing a good agreement for both the lower momentum (dominated by collision losses) 
and higher momentum (dominated by bremsstrahlung radiation) ranges.

\begin{figure}[t]
\centerline{
\epsfig{figure=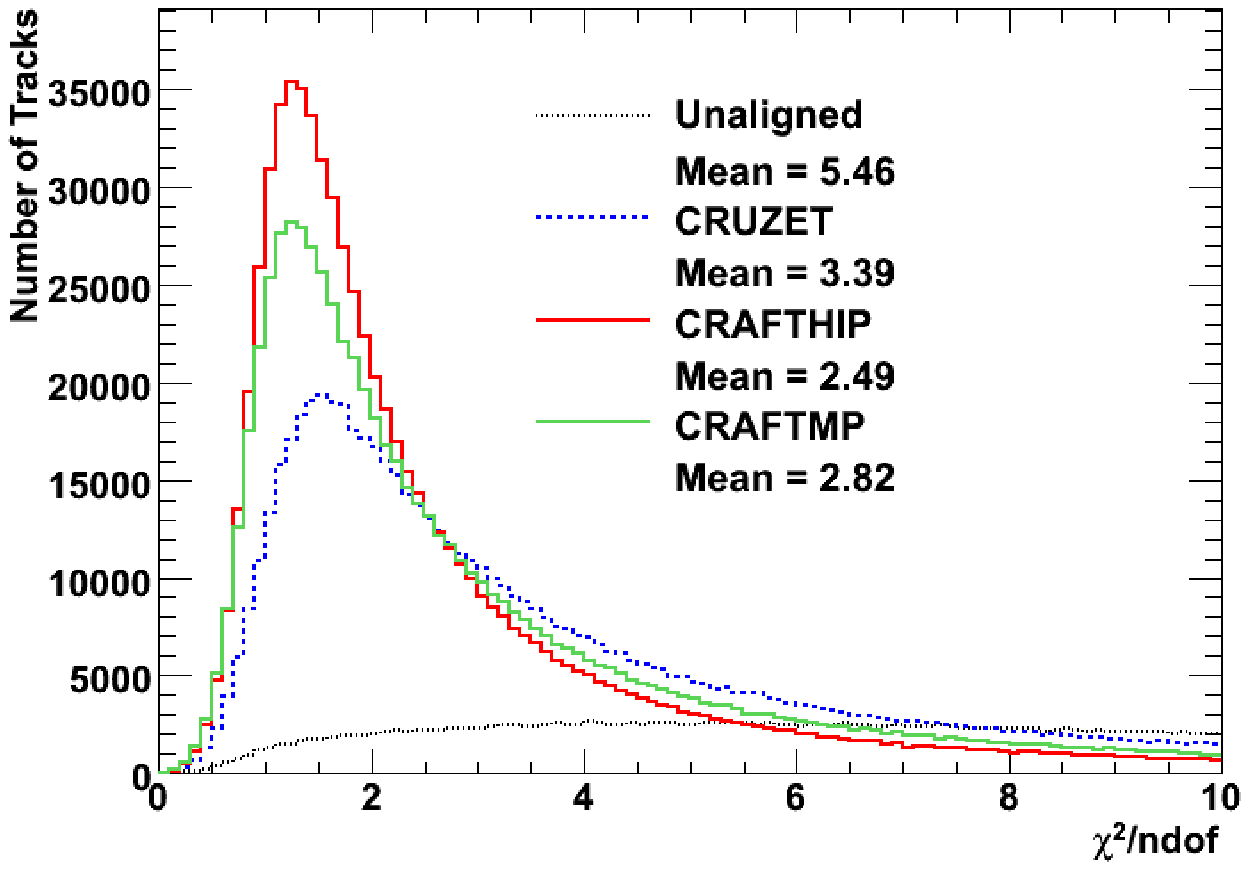,width=0.54\linewidth, angle=0}
\epsfig{figure=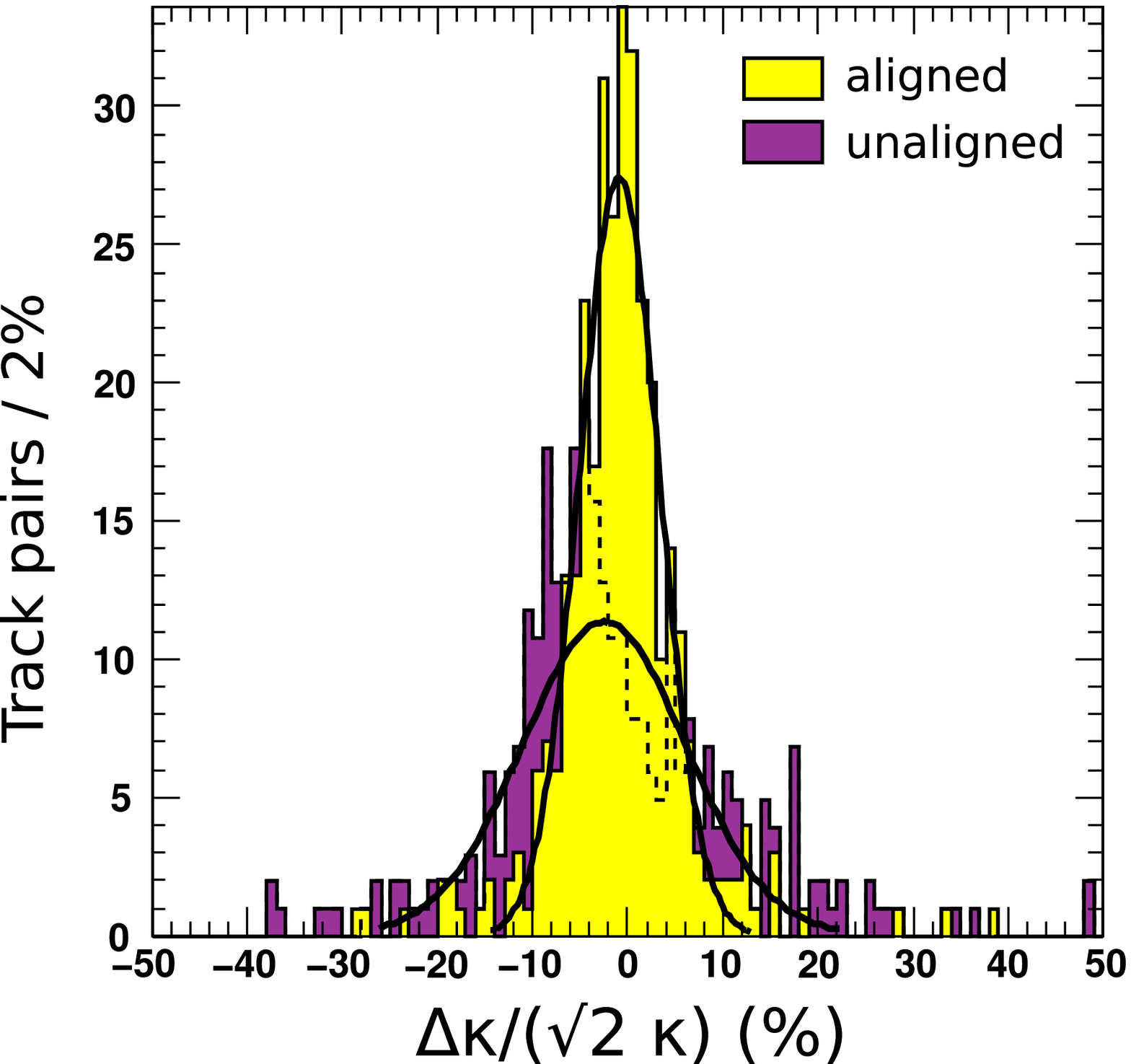,width=0.39\linewidth, angle=0}
}
\caption{(a): Improvement
in the quality of the reconstructed muon tracks with increase in the accuracy of tracker 
alignment measured using procedures based on {\it Millepede} and the {\it Hits 
and Impact Points} (HIP) algorithms. (b): Relative difference in curvature for tracks reconstructed using 
hits left by the same muon in either top or bottom part of the detector before 
(dashed) and after (solid) applying alignment corrections.
\label{fig:fig2}}
\end{figure}

Precise alignment of the tracking system is one of the most important pre-requisites for 
most physics analyses at the LHC. Cosmic ray data was used to measure actual positions
of sensitive elements of the tracker to obtain required alignment corrections~\cite{MP}. Figure~\ref{fig:fig2}(a) 
shows the improvement in the quality of muon tracks reconstructed in the tracking system as a
result of these studies. For very high $p_T$ muons, maintaining the good momentum resolution requires 
precise alignment of the muon system. Figure~\ref{fig:fig2}(b) shows the improvement in the accuracy 
of muon momentum reconstruction for higher momentum ($p_T>200$ GeV/c) muons with better precision of 
muon alignment. The plot compares curvature measured separately for the top and bottom part of the 
trajectory of a muon traversing the detector. Because the two tracks are reconstructed using
different sets of hits, this comparison provides an unbiased measurement of the muon momentum resolution.

\section{Early Physics Plans}

\begin{figure}
\centerline{
\epsfig{figure=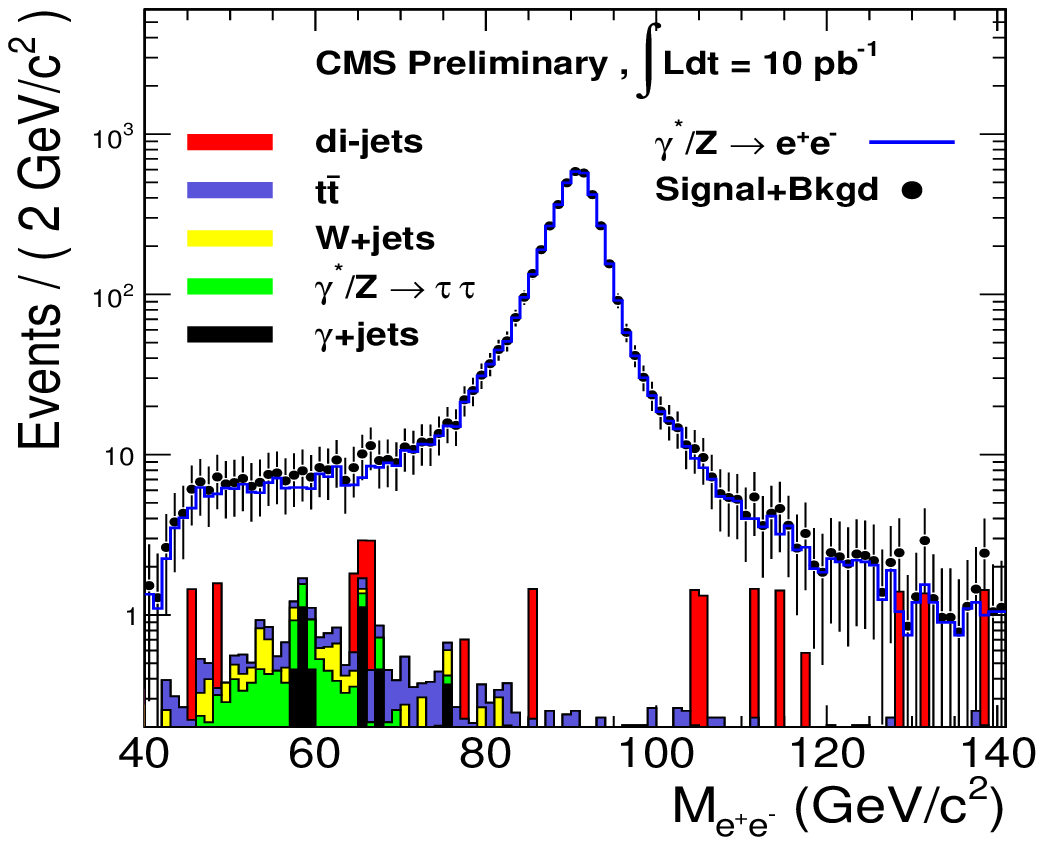,width=0.51\linewidth, angle=0}
\epsfig{figure=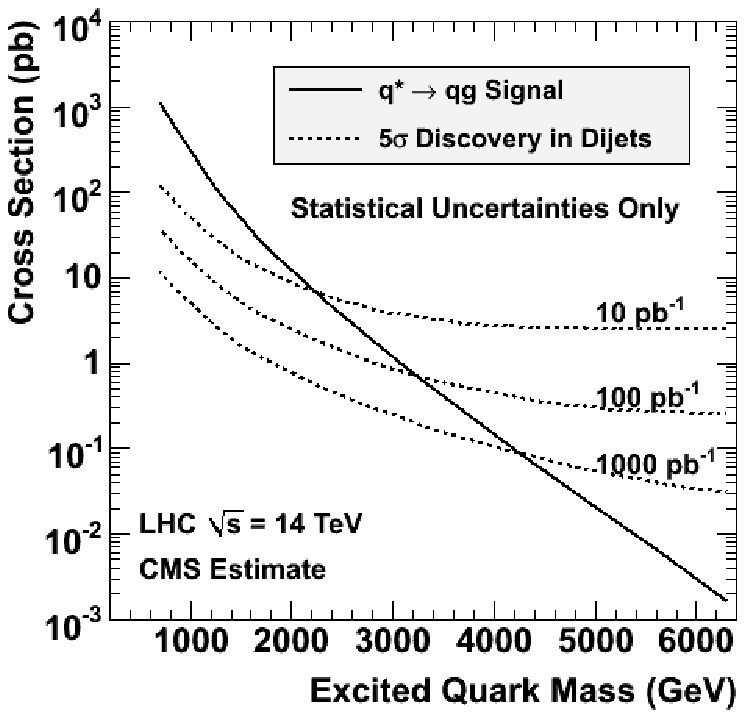,width=0.45\linewidth, angle=0}
}
\caption{(a): Expected di-electron invariant mass distribution for events 
selected in the $Z\to ee$ cross-section analysis with 10 pb$^{-1}$ of data 
at $\sqrt{s}=10$ TeV. (b): Reach for the excited quark production in early CMS data for 10-1000 pb$^{-1}$ of data.
\label{fig:fig3}}
\end{figure}

Ultimate certification of the detector performance can only be established using collisions data.
As with any other detector, the first physics measurements at CMS will aim at re-establishing
the SM ``standard candles'', tuning momentum and energy scales, understanding detector
effects and backgrounds. For example, measurement of the di-lepton signatures $J/\psi (\Upsilon)  \to \mu \mu$,
$Z \to ee (\mu \mu)$ will be critical in achieving precision calibration of tracking momentum and calorimeter 
energy scales. Figure~\ref{fig:fig3}-a shows the expected di-electron invariant mass distribution for events 
to be selected for the $Z\to ee$ cross-section analysis with just 10 pb$^{-1}$ of data at $\sqrt{s}=10$ TeV (at lower
$\sqrt{s}$, the number of expected events decreases but remains substantial). Apart from providing a test 
of higher order QCD calculations and parton distribution functions, this measurement will be critical in
tuning the absolute scale of the electromagnetic calorimeter and electron reconstruction techniques.

Similarly, measurements of $W \to e \nu$ and $W \to \mu \nu$ will be essential in understanding the 
missing energy resolution, which will be critical for future SUSY searches. The measurement of $Z \to \tau \tau$ 
cross-section will serve as an important test of cross-detector capabilities and understanding of tracking, 
calorimetry, validation of electron, muon, and hadronic tau reconstruction techniques, and the transverse 
missing energy resolution. At the LHC, an important new ``standard candle'' will be the $t\bar{t}$ events 
that will be produced in abundance due to a large increase in the production cross-section compared to the 
Tevatron. Experience in studying top properties will also be an invaluable tool in future searches in the 
context of SUSY and other new physics models.

While re-establishing the SM candles will have to precede any analyses targeting new physics, certain
searches depending on a fraction of detector sub-systems can be performed with already early data. One 
such example is a search for new resonances (or compositeness) in the di-jet channel. Figure~\ref{fig:fig3}-b 
shows the reach for excited quark in early CMS data showing sensitivity of searches with 10-1000 pb$^{-1}$ of data 
(results shown are for $\sqrt{s}=14$ TeV, for reference the excluded cross-section at $\sqrt{s}=10$ TeV is 
about 25-30\% higher).  Searches for other high $p_T$ signatures such as $Z^\prime$ and $W^\prime$ 
resonances in lepton modes will also surpass Tevatron sensitivity with already moderate amounts of
early data.

\section{Summary}

While the most recent delay in the LHC start has been unfortunate, the CMS experiment 
has used this time to make large gains in understanding detector calibration, alignment
and reconstruction techniques using real data collected during cosmic running. These
improvements and the invaluable experience of continuous long-term detector operation 
will help bring the first physics from the LHC sooner.

\section*{Acknowledgments}
The author would like to thank the organizers of the Blois conference for making this event
a true success. This contribution has been made possible due to the funding support provided 
by the US Department of Energy and the State of Texas.

\end{document}